\documentclass[twocolumn,pra,superscriptaddress]{revtex4-1}
\usepackage{amsfonts}
\usepackage{amssymb}
\usepackage{amsmath}
\usepackage{epsfig}
\usepackage{color}
\usepackage{graphics, graphicx}
\usepackage{bbold}
\usepackage{psfrag}
\usepackage{mathcomp}
\usepackage{subfigure}
\usepackage{verbatim}
\usepackage{float}
\usepackage{graphicx}
\usepackage[colorlinks,citecolor=blue]{hyperref}
\usepackage{natbib}
\makeatletter

\newcommand{\Rmnum}[1]{\expandafter\@slowromancap\romannumeral #1@}
\makeatother

\begin{document}
	\title{Detecting Complex-Energy Braiding Topology in a Dissipative Atomic Simulator with Transformer-Based Geometric Tomography}
	
	\author{Yang Yue}
	\thanks{These three authors contributed equally.}
	\affiliation{State Key Laboratory of Quantum Optics Technologies and Devices, Institute of Laser Spectroscopy, Shanxi University, Taiyuan 030006,
		China} 
	\affiliation{Collaborative Innovation Center of Extreme Optics,
		Shanxi University, Taiyuan 030006, China}

	\author{Nan Li}
	\thanks{These three authors contributed equally.}
	\affiliation{State Key Laboratory of Quantum Optics Technologies and Devices, Institute of Laser Spectroscopy, Shanxi University, Taiyuan 030006,
		China} 
	\affiliation{Collaborative Innovation Center of Extreme Optics,
		Shanxi University, Taiyuan 030006, China}	
	
     \author{Xin Zhang}
     	\thanks{These three authors contributed equally.}
	\affiliation{Institute of Intelligent Information Processing, Shanxi University, Taiyuan 030006,
		China} 
		
		\author{Chenhao Wang}
	\affiliation{State Key Laboratory of Quantum Optics Technologies and Devices, Institute of Laser Spectroscopy, Shanxi University, Taiyuan 030006,
		China} 
	\affiliation{Collaborative Innovation Center of Extreme Optics,
		Shanxi University, Taiyuan 030006, China}
		
				\author{Zeming Fang}
	\affiliation{State Key Laboratory of Quantum Optics Technologies and Devices, Institute of Laser Spectroscopy, Shanxi University, Taiyuan 030006,
		China} 
	\affiliation{Collaborative Innovation Center of Extreme Optics,
		Shanxi University, Taiyuan 030006, China}
		
		\author{Zhonghua Ji}
	\affiliation{State Key Laboratory of Quantum Optics Technologies and Devices, Institute of Laser Spectroscopy, Shanxi University, Taiyuan 030006,
		China} 
	\affiliation{Collaborative Innovation Center of Extreme Optics,
		Shanxi University, Taiyuan 030006, China}
			
	\author{Liantuan Xiao}
	\affiliation{State Key Laboratory of Quantum Optics Technologies and Devices, Institute of Laser Spectroscopy, Shanxi University, Taiyuan 030006,
		China} 
	\affiliation{Collaborative Innovation Center of Extreme Optics,
		Shanxi University, Taiyuan 030006, China}
	
	\author{Suotang Jia}
	\affiliation{State Key Laboratory of Quantum Optics Technologies and Devices, Institute of Laser Spectroscopy, Shanxi University, Taiyuan 030006,
		China} 
	\affiliation{Collaborative Innovation Center of Extreme Optics,
		Shanxi University, Taiyuan 030006, China}
				
		\author{Yanting Zhao}
	\thanks{zhaoyt@sxu.edu.cn}
	\affiliation{State Key Laboratory of Quantum Optics Technologies and Devices, Institute of Laser Spectroscopy, Shanxi University, Taiyuan 030006,
		China} 
	\affiliation{Collaborative Innovation Center of Extreme Optics,
		Shanxi University, Taiyuan 030006, China}
		
	\author{Liang Bai}
	\thanks{bailiang@sxu.edu.cn}
	\affiliation{Institute of Intelligent Information Processing, Shanxi University, Taiyuan 030006, China}

		 \author{Ying Hu}
	\thanks{huying@sxu.edu.cn}
	\affiliation{State Key Laboratory of Quantum Optics Technologies and Devices, Institute of Laser Spectroscopy, Shanxi University, Taiyuan 030006,
		China} 
	\affiliation{Collaborative Innovation Center of Extreme Optics,
		Shanxi University, Taiyuan 030006, China}

\begin{abstract}
Machine learning (ML) is shaping our exploration of topological matter, whose existence is inherently tied to the geometry of quantum states or energy spectra. In non-Hermitian systems, distinctive spectral geometry can lead to topological braiding of complex-energy bands, yet directly probing this topology-geometry interplay remains challenging. Here, we introduce a Transformer-based ML framework to capture this interplay and experimentally demonstrate it in a dissipative cold-atom simulator. Using a Bose-Einstein condensate, we engineer tunable dissipative two-level systems 
whose complex eigenenergies form braids. Owing to the density‑dependent dissipation, the instantaneous energy braids exhibit topologically distinct structures at short and long times. The Transformer not only accurately predicts topological invariants for diverse energy braids but also, through its self-attention mechanism, autonomously highlights band crossings as the governing underlying geometric feature. Our work paves the way for ML-guided exploration of non-Hermitian topological phases in cold atoms and beyond.
 \end{abstract}
\maketitle	

\noindent\textbf{Introduction} 

Machine learning (ML), an effective data-driven approach in artificial intelligence, is transforming the study of topological matter by enabling automated analysis of complex quantum systems~\cite{Carrasquilla2017,Carleo2019,Schuttet2020,Bhart2020,Wang2023,Lin2023}. The global topological character of topological phases originate from the geometric properties of their wave function and energy spectrum, giving rise to intriguing physics: nonlocal topological invariants classify these phases into distinct categories, and ensure robust properties such as quantized transport and non-Abelian quasiparticles, with potential applications from semiconductor spintronics to quantum computation~\cite{Hasan:RMP2010,Qi:RMP2011,Chiu:RMP2016,Lu:NP2014}. Recent advances demonstrate that supervised neural networks, such as convolutional neural networks (CNNs), can learn topological invariants from local data~\cite{Zhang2018,Cong2019,Rem2019,Lian2019,Zhao2022}. Unsupervised learning approaches such as diffusion maps can classify diverse topological phases without data label~\cite{Chen2024,Rodriguez2019,Scheurer2020,Lustig2020,Yu2021,Yu2022,Long2024,Lustig2020,Kaming2021}, ranging from symmetry-protected band insulators to disordered photonic materials, even when the classification is not \textit{a priori} known. Moreover, ML facilitates direct detection of topological phase transitions~\cite{Rem2019,Lian2019,Lustig2020,Kaming2021,Yu2022,Zhao2022} from experimental data, including single-shot images that defy conventional analysis.

\begin{figure*}[tb]
 \begin{center}  
\includegraphics[width=1\textwidth]{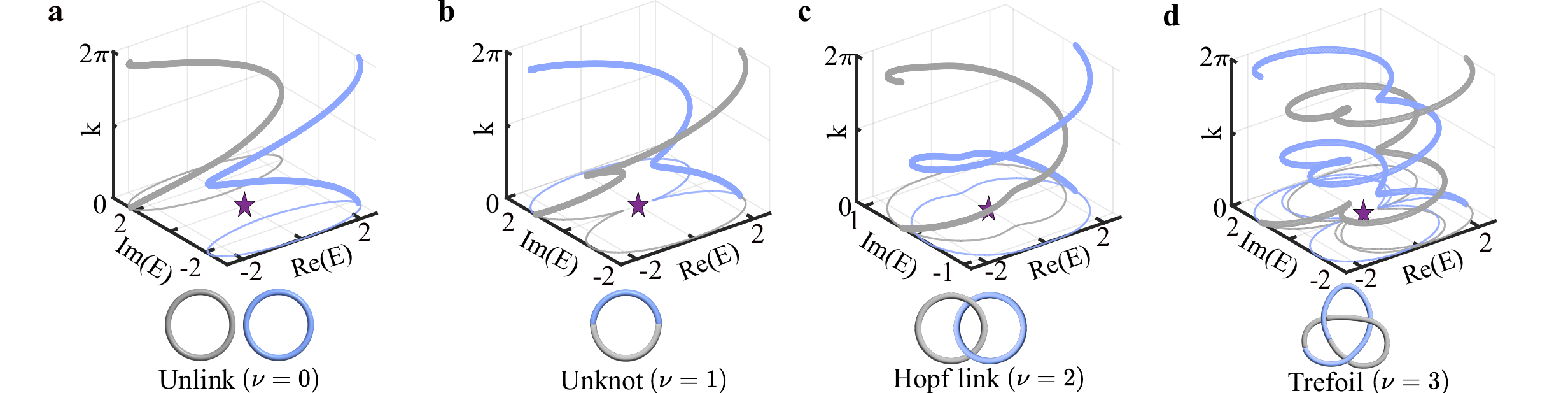}  
 \caption{\textbf{Illustration of topological complex-energy braiding in energy-momentum space.} The gray and blue curves denote two bands, respectively, and the star labels the exceptional point. Gluing the $k = 0$ and $k = 2\pi$ planes leads to different knotted structures, such as $\mathbf{a}$ the unlink, $\mathbf{b}$ the unknot, $\mathbf{c}$ the Hopf link, and $\mathbf{d}$ the Trefoil knot, distinguished by braid degree $\nu=0,1,2,3$, respectively. In all plots, theoretical calculations are based on Eq.~(\ref{eq1}) with $J_1 = 0.7$, $J_2 = 1.8$, $J_3 = 1$, and $\Omega = 0.5$. For parameters ($\gamma$,m), we use $\mathbf{a}$ $(4.5,1)$, $\mathbf{b}$ $(2, 1)$, $\mathbf{c}$ $(0.5,1)$, and $\mathbf{d}$ $(2,3)$.}  \label{Fig1}
 \end{center} 
\end{figure*}

\begin{figure*}[tb] 
 \begin{center} 
	\includegraphics[width=1\textwidth]{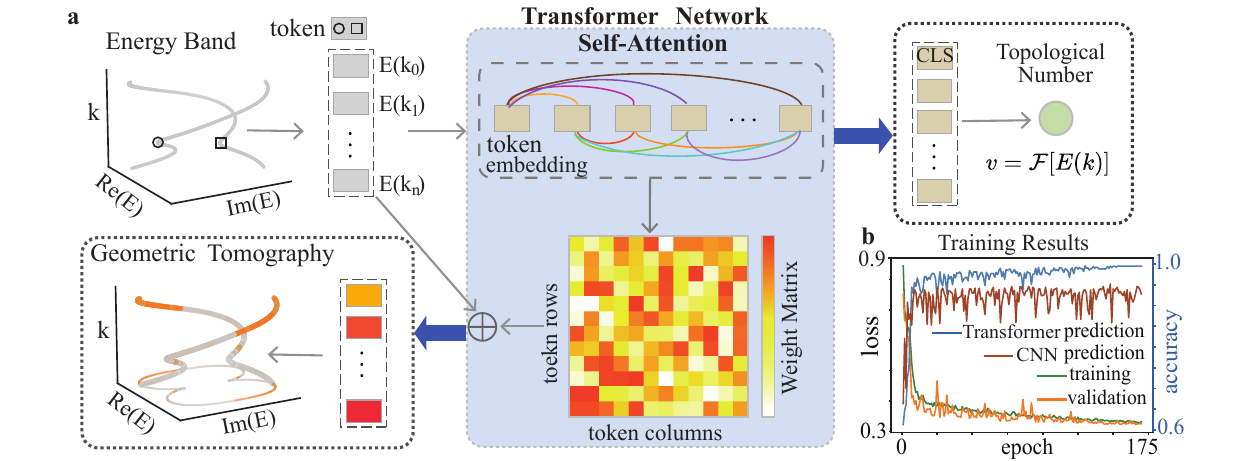}  
	\caption{\textbf{Joint learning of complex-energy braiding topology and geometry via the self-attention of the Transformer.} $\mathbf{a}$ Schematics of the working principle. The complex-energy bands in the momentum space ($k\in [0,2\pi]$) are encoded as sequential input data, where the two eigenenergies $E_\pm(k)$ at each $k$-point correspond to a feature token. The Transformer processes this sequence and outputs the topological invariant $\nu$ as a functional mapping of the input energy bands. Leveraging the unique self-attention mechanism (blue shaded region) of the Transformer, the model generates attention weight matrices that, when projected into the input data, reveal its focus on distinct regions of energy bands in classifying the bands' braiding topology. The primary focus of attention locates the decisive moment-energy points underlying braiding topology. $\mathbf{b}$ Training results. Blue curves depict the prediction accuracy of Transformer on the test dataset. The green (yellow) curves show training (validation) loss. As a comparison, the prediction accuracy of CNN trained under identical conditions is shown by the brown curve.} \label{Fig2}
 \end{center} 
\end{figure*}

Despite these successes, however, it remains challenging to \textit{simultaneously} identify topological invariants of phases \textit{and} reveal their underlying geometric origins for conventional ML approaches. For instance, CNNs' black-box nature often precludes direct geometric interpretation of predicted topology, particularly without manual feature engineering. Similarly, unsupervised methods typically cluster phases based on data similarities and connections, rather than learning physically meaningful geometric features in individual samples, thereby limiting deeper insights from their predictions.

Yet achieving this dual capability is essential for understanding and predicting observable topological phenomena. On the one hand, the very existence of topological phases is intrinsically tied to the geometry of eigenstates and energy spectrum. A paradigmatic example is in quantum Hall systems, where Berry curvature, a geometric property characterizing the phase variation of eigenstates across the Brillouin zone (BZ), integrates to yield the topological invariant (Chern number) that governs quantized Hall response~\cite{Xiao:RMP2010}. Recent development of quantum geometry~\cite{Torma2023,Liu2025} further illuminates how the metric of eigenstates affects topological transport. Meanwhile, burgeoning research in open (non-Hermitian) systems has uncovered new topological phenomena, such as topological complex-energy braiding, in which the geometry of energy spectrum provides the backbone~\cite{Shen2018,Bergholtz2021,Ding2022,Wang2021,Patil2022,Zhang2023,Cao2023,Wu2023,Rao2024}. On the other hand, in cold-atom experiments, directly probing both topological invariant and its geometric origins remains a challenge: the former is inaccessible to local measurements; while the latter traditionally relies on indirect, sophisticated methods such as momentum-resolved interferometry or quantum state tomography~\cite{Hauke2014,Flaschner2016} - especially demanding in dissipative settings~\cite{Li2019,Ren2022,Wang2024}.

\begin{figure*}[tp]
 \begin{center}  
\includegraphics[width=1\textwidth]{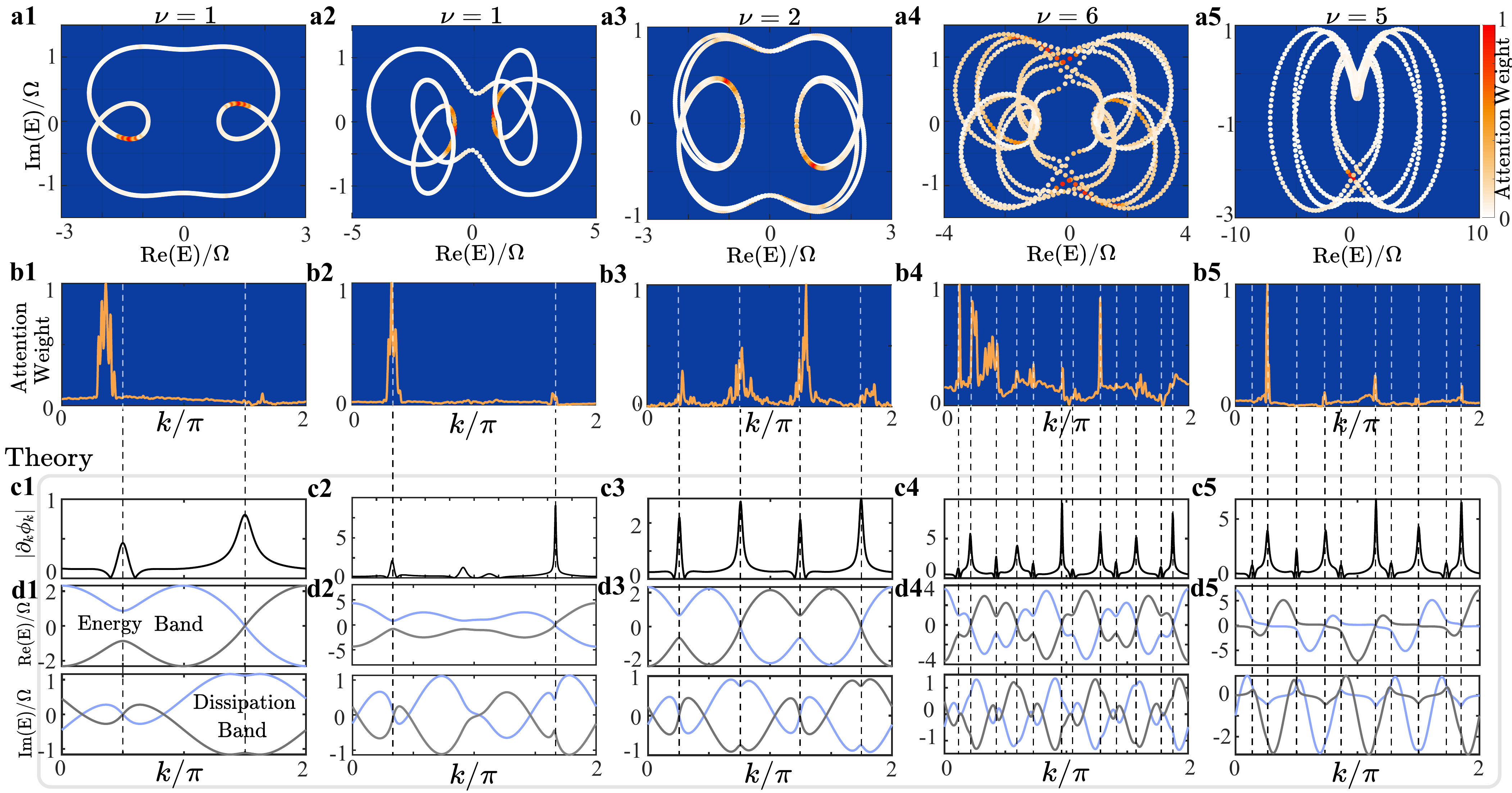}  
 \caption{\textbf{Attention-based geometric tomography of complex-energy bands.} Attention weights generated from the Transformer are projected onto the test data. $\mathbf{a1}$-$\mathbf{a5}$ Distribution of attention weights shown in the complex energy plane, for energy braids with distinct braiding degree $\nu$. $\mathbf{b1}$-$\mathbf{b5}$ Corresponding distribution of attention weights in the Brillouin zone (BZ). \color{black}$\mathbf{c1}$-$\mathbf{c5}$ Theoretically calculated distribution of $|\partial_k\phi_k|$ in momentum space (see text). $\mathbf{d1}$-$\mathbf{d5}$ Real and imaginary parts of the complex energy bands. Dashed line denotes band crossing with extreme $|\partial_k\phi_k|$ in the real or imaginary component of energy bands. Test data are generated from Eq.~(\ref{Etest}) with $\Omega=1$; other parameters $(\alpha, J_1,J_2,J_3,J_4,\gamma,m)$ are $\mathbf{a1}$-$\mathbf{d1}$ $(0,0.1,2,1,0,1,1)$; $\mathbf{a2}$-$\mathbf{d2}$ $(0,2,2,1,0.2,0.5,2)$; $\mathbf{a3}$-$\mathbf{d3}$ $(0,0.1,2,1,0,0.5,2)$; $\mathbf{a4}$-$\mathbf{d4}$ $(0,1.5,2,1,0,1,6)$; $\mathbf{a5}$-$\mathbf{d5}$ $(1,1.5,2,1,0,1,5)$. \color{black}}  \label{Fig3}
 \end{center} 
\end{figure*}

To address the aforementioned challenges, in this work we present an approach based on the state-of-the-art ML technique, the Transformer~\cite{Vaswani2017}. Unlike CNNs or diffusion map methods, the Transformer leverages its self-attention mechanism to compute pairwise correlations across samples, which allows to simultaneously identify topological invariant and visualize key spots for phase recognition. We show how a suitably trained Transformer explicitly captures the geometry-topology connection in complex-energy braids of non-Hermitian bands, ranging from links to quite intricate knots. 

In particular, {we experimentally construct topological complex-energy braids in an atomic simulator and demonstrate their characterization using the Transformer}. By engineering a tunable two-level dissipative system with Bose-Einstein condensates (BEC) of $^{87}\textrm{Rb}$ atoms, we measure their eigenvalues that form distinct braids.  Different from the single-particle dissipation implemented in previous atomic experiments~\cite{Li2019,Ren2022} and trapped-ion simulations~\cite{Cao2023}, the dissipation engineered in a BEC exhibits density dependence due to the collective atomic response to resonant light~\cite{You1994,Pellegrino2014,Jenkins2016,Schilder2020,Wang2024}. This results in dynamically evolving energy braids whose instantaneous structures in the short- and long-time limits can be topologically distinct. \color{black}We find the Transformer, trained solely on simulated symmetric energy bands from a single-particle Hamiltonian, generalizes effectively to experiment. Not only does it accurately predict the topological invariants for various energy braids, but, crucially, it autonomously identifies band crossings, namely intersections in the real or imaginary parts of the complex-energy band\color{black}, as the governing geometric features through its attention weight. Our work opens a new path to explore exotic topological phases in open quantum systems. It also paves a way for exploring fascinating interplay between quantum geometry and topology, where experimental studies remain scarce and challenging~\cite{Tan2019,Cuerda2024,Kim2025}. 

 \color{black}

\bigskip
\noindent\textbf{Results} 

\noindent\textbf{Complex-Energy Braiding Topology} 

We first summarize the key aspect of complex-energy braiding topology, taking the example of a Bloch Hamiltonian with sublattice symmetry $
H(k) = (1/2)(\Delta_k - i\Gamma_k)\sigma_z+\Omega\sigma_x$. Here, $k\in [0,2\pi]$ is quasimomentum, $\sigma_{x,z}$ are Pauli matrices, $g$ is the coupling strength, $\Delta_{k}=2J_{1}\sin(k)+2J_4\cos(2k)+2J_{2}\cos(mk)$ and $\Gamma_{k}=-2J_{3}\sin(mk)+\gamma$, with real parameters $\Omega$ and $J_{1,2,3}$. The $H(k)$ exhibits two separated energy bands 
\begin{equation}\label{eq1}
{E}_{\pm}(k)= \pm \sqrt{({\Delta_k - i\Gamma_k})^2/4 + \Omega^2},
\end{equation}
which are symmetric around $E=0$ due to the sublattice symmetry $\sigma_zH(k)\sigma_z=-H(k)$. Since ${E}_{\pm}(k)$ is complex-valued, their paths wind in the complex energy plane as $k$ varies across BZ. Owing to periodicity $E_\pm(k)=E_\pm(k+2\pi)$, trajectories of ${E}_{\pm}(k)$ in the 3D energy-momentum ($E$-$k$) space thus braid into knots or links~\cite{Wang2021}, such as the unlink [Fig.~\ref{Fig1}$\mathbf{a}$], the unknot [Fig.~\ref{Fig1}$\mathbf{b}$], the Hopf link [Fig.~\ref{Fig1}$\mathbf{c}$], the Trefoil knot [Fig.~\ref{Fig1}$\mathbf{d}$], and others. 

The topology of distinct two-band braids can be classified by the braid group $\mathbb{B}_2$ associated with an integer topological invariant - the braid degree. Let $\tilde{E}_{\pm}(k)={E}_{\pm}(k)-E_0(k)$ with $E_0(k)=\textrm{Tr}(H(k))/2$, the braid degree is defined as
\begin{equation}\label{eq2}
\nu = \int_{0}^{2\pi} \frac{\mathrm{d}k}{2\pi i} \left( \partial_k \ln \tilde{E}_{+}(k) + \partial_k \ln \tilde{E}_{-}(k) \right),
\end{equation}
which is related to the phase winding $\partial_k\phi_\sigma(k)$ of $E_\sigma(k)=|E_\sigma(k)|e^{i\phi_\sigma(k)}$ ($\sigma=\pm$)~\cite{Shen2018}. A nonzero $\nu$ indicates that the spectral loops in the complex energy plane encircles an exceptional point (EP), where $E_+(k)=E_-(k)$ (star in Fig.~\ref{Fig1}), and the value of $\nu$ counts how many times the two bands braid in the $E$-$k$ space as $k$ traverses the BZ. For instance, different braids in Fig.~\ref{Fig1} correspond to $\nu=0,1,2,3$, respectively.
\bigskip

\noindent\textbf{Attention-enabled joint learning of braiding topology and geometry}
\begin{figure*}[t]
 \begin{center}  
\includegraphics[width=1\textwidth]{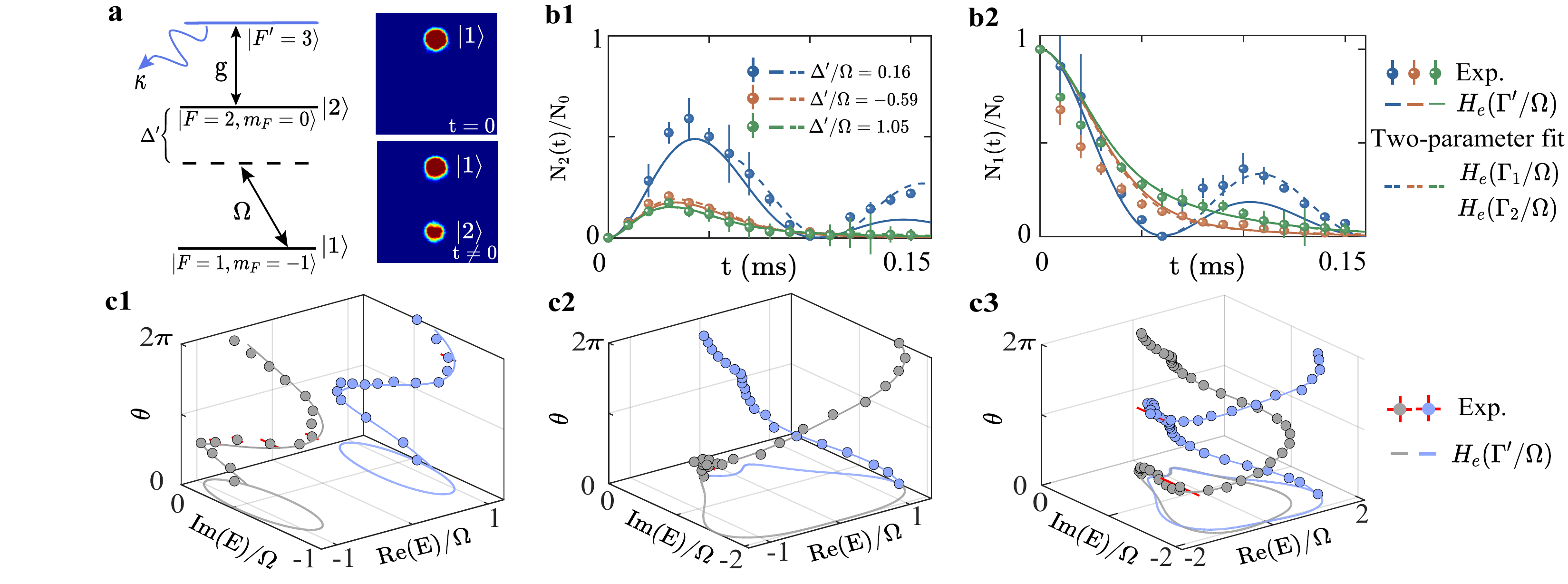}  
 \caption{\textbf{Experimental setup with atomic BECs.} $\mathbf{a}$ Level diagram. A microwave field detuned by $\Delta$ coherently couples the hyperfine states $|F=1,m_F=-1\rangle$ and $|F=2,m_F=0\rangle$ of $^{87}\textrm{Rb}$ atoms with a coupling rate $\Omega$. A resonant optical beam is used to generate dissipation (atom loss) in the $|F=2,m_F=0\rangle$ state. The BEC is initialized in $|F=1,m_F=-1\rangle$. After an evolution time $t$, the Stern-Gerlach absorption image is taken after $10$ ms time of flight. $\mathbf{b1(b2)}$ Measured $N_{2(1)}(t)/N_0$ as a function of time, where $N_0$ is initial atom number. The blue, yellow and green dots denote experimental data under various [$\Delta^{\prime}/\Omega, \Gamma'/\Omega$]: $[0.16,0.53]$ when $m=1$ (blue), $[-0.59,1.60]$ when $m=1$ (yellow), and $[1.05,1.71]$ when $m=2$ (green), respectively. Each experimental data is the average over 3 measurements. The error bars are $1\sigma$ standard deviation. Solid curves depict simulations via Hamiltonian $H_e$ in Eq.~(\ref{eq4}) with $\Gamma'$. Dashed curves denote two-parameter fit to the data using $H_e(\Gamma_1)$ and $H_e(\Gamma_2)$; see Methods and Supplementary Sec.~III. \color{black}$\mathbf{c1}$-$\mathbf{c3}$ Measured energies in the short-time regime, shown in ($\textrm{Re}E$, $\textrm{Im}E$, $\theta$) space. Experimental data of eigenvalues are extracted from the fitted $H_e(\Gamma_1)$ (Methods). Solid curves denote calculated eigenvalues of $H_e(\Gamma')$; parameters $(J_1,J_2,J_3,\gamma,m)$ are $\mathbf{c1}$ $(0,1/4,1/4,1,1)$; $\mathbf{c2}$ $(0,1/2,1/4,2,1)$ and $\mathbf{c3}$ $(1/8,1/2,1/4,2,2)$. In all measurements, $\Omega=31.4$ kHz.\color{black}}  \label{Fig4}
 \end{center} 
\end{figure*}

We train a Transformer to jointly learn braiding topology and its geometric underpinning, as schematically illustrated in Fig.~\ref{Fig2}$\mathbf{a}$. The architecture details can be found in Methods and Supplementary Sec.~I and Sec.~II. The model is trained exclusively on symmetric spectra $E_\pm(k)$ simulated from Eq.~(\ref{eq1}). The BZ is discretized into $N_k=500$ uniformly spaced $k$-points to ensure sufficient momentum resolution. The input data consists of a $N_k \times 4$ matrix representing the two complex-energy bands; each row (i.e., token) encodes the two complex eigenvalues: $(\textrm{Re}E_+(k),\textrm{Im}E_+(k),\textrm{Re}E_-(k),\textrm{Im}E_-(k))$ at momentum $k$. These energy data is fed into an embedding layer, which adds a learnable classification (CLS) token, whose feature dimensionality is expanded from 4 to 16 dimensions. Then, the Transformer encoder employs multi-head self-attention mechanisms to compute pairwise correlations across all $k$-points, enabling the CLS token to learn features in the energy bands comprehensively. These learned features are ultimately converted to probabilities corresponding to different braiding degrees $\nu$. Concurrently, the attention weight in the input data visibly highlights the key tokens essential for learning $\nu$. 

\begin{figure*}[tp]
	\begin{center} 
		\includegraphics[width=1\textwidth]{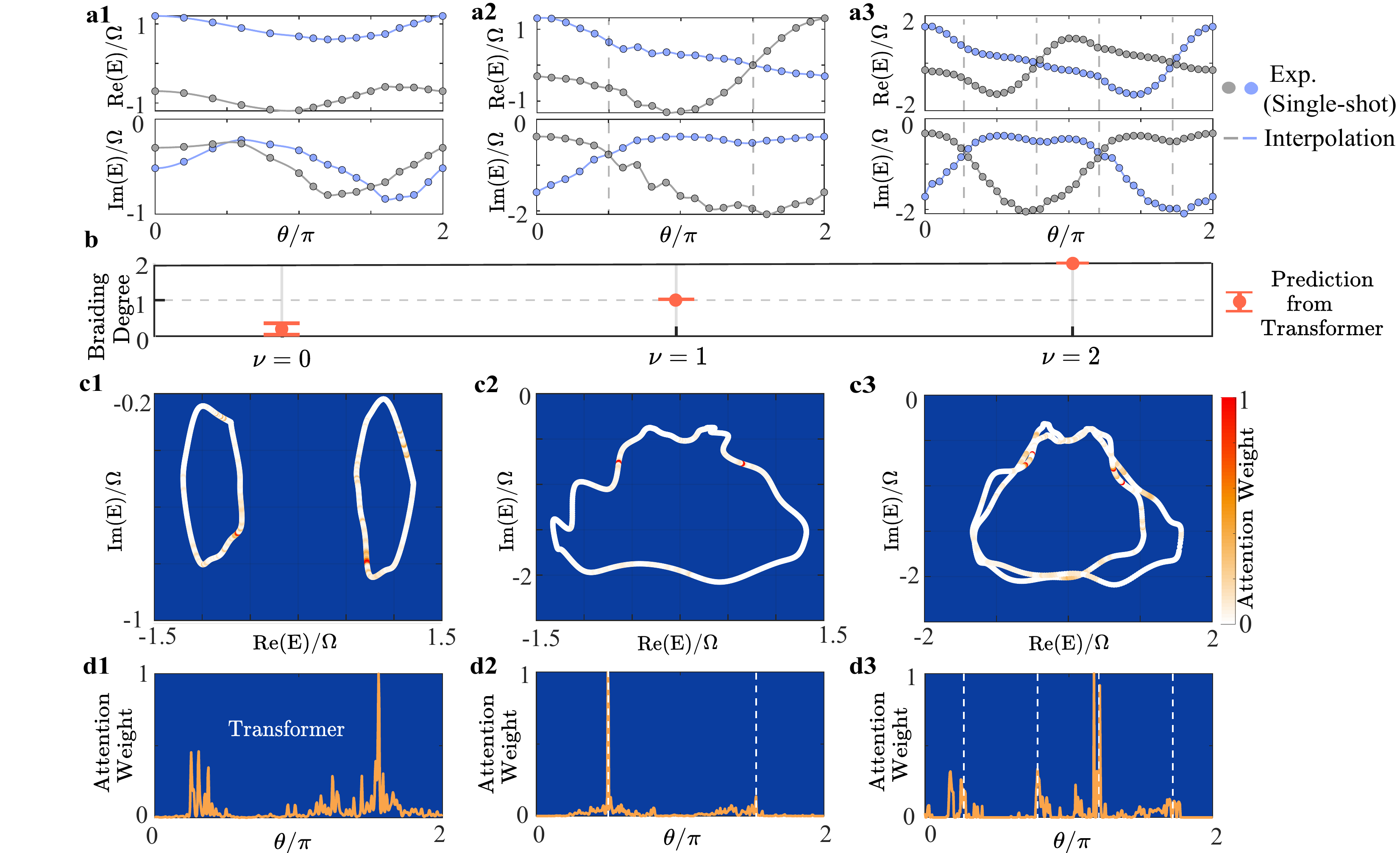}  
		\caption{\textbf{Experimental detection of topological and geometrical properties of energy braids in the short-time regime via a trained Transformer.} $\mathbf{a1}$-$\mathbf{a3}$ Single-shot measurement of complex energy $E$ as a function of $\theta\in[0,2\pi]$. The plots correspond to Figs.~\ref{Fig4}$\mathbf{c1}$-$\mathbf{c3}$, respectively, where experimental data denote eigenvalues of $H_e(\Gamma_1)$. Solid lines are the interpolated curve passing through experimental data (see Methods and Supplementary Sec.~IV). $\mathbf{b}$ Transformer-predicted braiding degree $\nu$. For each set of single-shot measurement of complex eigenvalues, braiding degree $\nu$ is extracted using the Transformer. The average $\nu$ from three set of measurements is shown, and the error bar denotes standard deviation. $\mathbf{c1}$--$\mathbf{d3}$ Attention weights projected onto the preprocessed experimental data, shown in $\mathbf{c1}$-$\mathbf{c3}$ in complex energy plane, and $\mathbf{d1}$-$\mathbf{d3}$ as a function of $\theta\in [0,2\pi]$. In the top and bottom rows, dashed line denotes band crossings with extreme phase gradients. }\label{Fig5}
	\end{center} 
\end{figure*}

The self-attention mechanism of the Transformer (Methods) fundamentally differs from CNN in capturing the nonlocal correlations essential for topological classification [Fig.~\ref{Fig2}$\mathbf{a}$]. While CNNs rely on localized receptive fields, self-attention operates on the complete set of $k$-points at once, enabling direct access to both topological invariant and geometric features - information often obscured in traditional convolutional architectures. Specifically, the self-attention mechanism computes attention weights that quantify the influence of each $k$-point on every other point within the BZ, and dynamically re-weight spectral features during classification. As shown below, this allows the model to autonomously identify topologically relevant band regions without requiring \textit{a priori} feature engineering.

Following standard training protocols, we split data into training (80\%) and validation (20\%) sets to ensure robust evaluation of generalization performance. The model is optimized via stochastic gradient descent to minimize cross-entropy loss, converging in 170 epochs, as evidenced by stabilized validation metrics in Fig.~\ref{Fig2}$\mathbf{b}$. 

We test the trained Transformer with unseen data (see Methods), including those in the absence of sublattice symmetry. The test results are summarized below:

(1) The model demonstrates exceptional performance in predicting braiding degree $\nu$, achieving 99.93\% classification accuracy [Fig.~\ref{Fig2}$\mathbf{b}$], substantially outperforming CNN under identical conditions (Supplementary Sec.~I).  

(2) The attention heatmap in Fig.~\ref{Fig3} indicates that band crossing is the geometric feature directly relevant to braiding topology. Figures~\ref{Fig3}$\mathbf{a1}$-$\mathbf{b5}$ present the heatmap associated with various $\nu$, shown in the complex energy plane and momentum space, respectively. To see their physical implications, we calculate $|\partial_k\phi_k|$ with $\phi_k=\phi_+(k)+\phi_-(k)$ in the BZ [Figs.~\ref{Fig3}$\mathbf{c1}$-$\mathbf{c5}$] and the complex-energy band structures [Figs.~\ref{Fig3}$\mathbf{d1}$-$\mathbf{d5}$]. We find the attention peak in the momentum space [Figs.~\ref{Fig3}$\mathbf{b1}$-$\mathbf{b5}$] consistently aligns with band crossings in the real-energy or dissipation-rate bands, even for quite complex braiding patterns. Strikingly, when there are several crossings, as exemplified by Figs.~\ref{Fig3}$\mathbf{d1}$-$\mathbf{d4}$, attention weights correctly highlight those crossings with extreme phase gradient $|\partial_k\phi_k|$ as being topologically relevant. This occurs even for braiding degree as high as $\nu=6$ (Fig.~\ref{Fig3}$\mathbf{a4}$-$\mathbf{d4}$), when the energy structure becomes quite complicated. This selective attention focusing demonstrates that the Transformer learns not just global topology but also physically significant spots, thus guiding a geometric interpretation. 

(3) Equally remarkably, although the model is trained solely on symmetric spectra of the form in Eq.~(\ref{eq1}), it generalizes well to (simulated) asymmetric spectra (Methods), as illustrated by Figs.~\ref{Fig3}$\mathbf{a5}$-$\mathbf{d5}$. This indicates that the trained Transformer learns that sublattice symmetry is irrelevant for $\nu$, as one would expect from the formula~(\ref{eq1}). 

\bigskip

\noindent\textbf{Experimental demonstration in dissipative BEC} 

Equipped with the trained Transformer, we turn to the experimental study of complex-energy braiding based on the $^{87}\textrm{Rb}$ BEC with atomic density $n\sim 5.14\times 10^{13}\textrm{cm}^{-3}$. As shown in Fig.~\ref{Fig4}$\mathbf{a}$, two ground-state hyperfine states of $^{87}\textrm{Rb}$ atoms, $|1\rangle=|F = 1,m_{F}=-1\rangle$ and $|2\rangle=|F = 2,m_{F}=0\rangle$ is coupled by a microwave (MW) field detuned by $\Delta'$, with strength $\Omega$. Natural atomic loss is negligible over the relevant time scales. To introduce tunable dissipation on state $|2\rangle$, we use a laser light at $\lambda=780$ nm to drive a resonant transition from $|2\rangle$ and an electronically excited state $|F' = 3\rangle$ with linewidth $\kappa=2\pi\times 6.06$ MHz, resulting in a loss rate $\Gamma'$ of atomic populations in $|2\rangle$. \color{black}We have checked that the dissipation laser beam causes no visible resonance shift in $|2\rangle$. The setup effectively implements the Hamiltonian 
\begin{equation}\label{eq4}
 H_e= \left( \frac{\Delta' - i\Gamma'}{2} \right) I - \left( \frac{\Delta' - i\Gamma'}{2} \right) \sigma_z + \Omega \sigma_x. 
\end{equation}
 In our experiment, we fix $\Omega=31.4$ kHz, while independently vary $\Delta'$ (by tuning the MW frequency) and $\Gamma'$ (by tuning the laser power), respectively. \color{black}These parameters are sampled according to the relation $\Delta^{\prime}/\Omega=2J_1 \cos (\theta)+2J_2\cos (m\theta)$ and $\Gamma'/\Omega= -2J_3\sin (m\theta)+ \gamma$ with the integer $m$ and $\theta\in [0,2\pi]$; see Methods. By choosing suitable coefficients $J_1,J_2,J_3,\gamma$ and $m$, we realize the Hamiltonian $H_e(\theta)$ that hosts desired energy braids. \color{black}
Since Eq.~(\ref{eq4}) explicitly breaks sublattice symmetry, the eigenvalues
\begin{equation}\label{Etheta}
E_\pm (\theta)=\frac{\Delta' - i\Gamma'}{2}\pm \delta;\hspace{2mm}\delta=\sqrt{ \left( \frac{\Delta' - i\Gamma'}{2} \right)^2+\Omega^2}.
\end{equation}
form two asymmetric bands when $\theta$ varies in $[0,2\pi]$, in contrast to the symmetric spectra~(\ref{eq1}). 

\begin{figure}[tp]
	\begin{center} 
		\includegraphics[width=1\columnwidth]{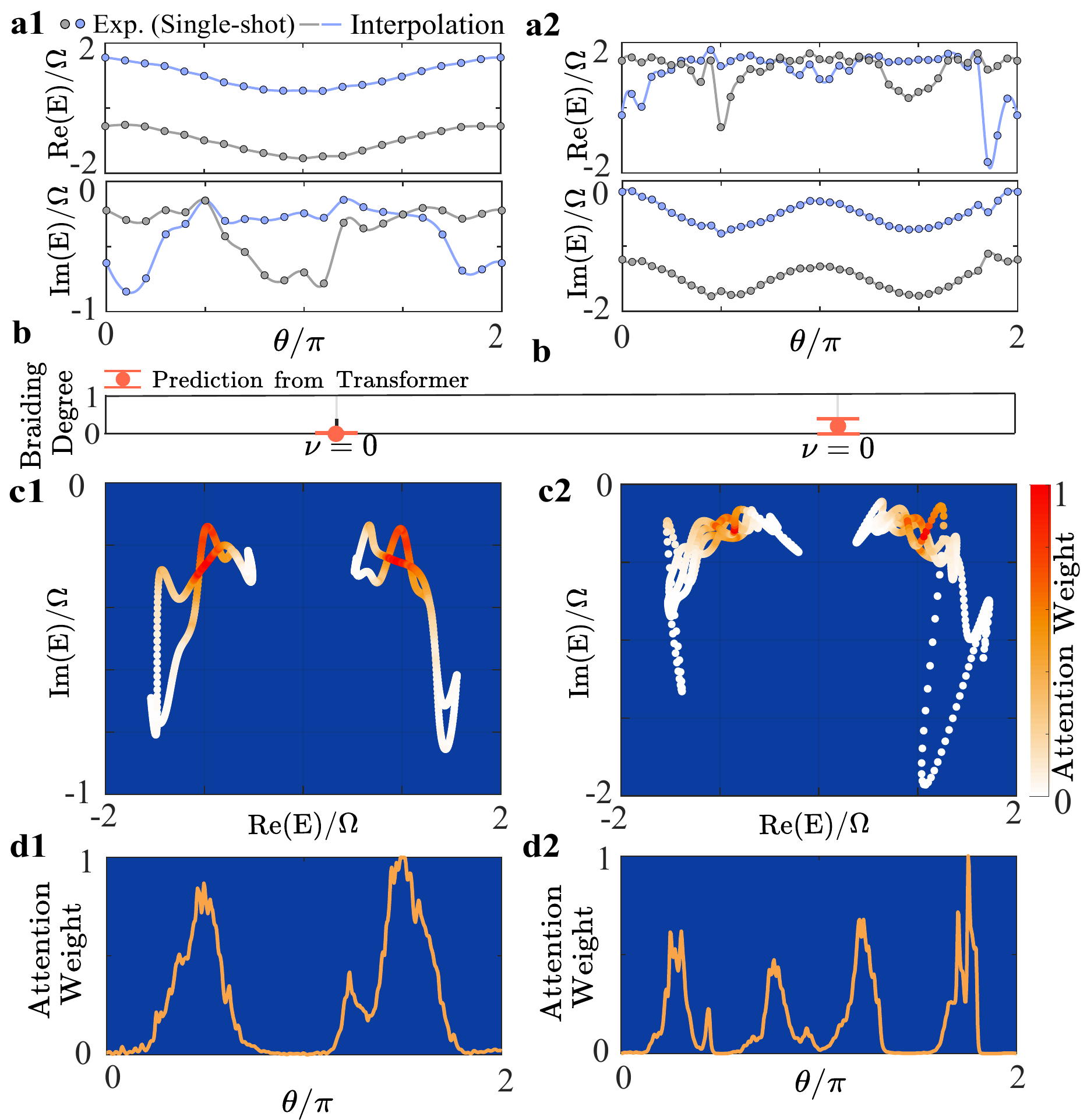}  
		\caption{\textbf{Long-time energy braids corresponding to the $\mathbf{a1}$ unknot and $\mathbf{a2}$ Hopf link shown in Fig.~\ref{Fig5}.}  $\mathbf{a1}$-$\mathbf{a2}$ Long-time counterpart for the spectra in Figs.~\ref{Fig5}$\mathbf{a2}$-$\mathbf{a3}$, respectively. Experimental data are eigenvalues obtained from $H_e(\Gamma_2)$; solid curve is the interpolation. $\mathbf{b}$ Braid degree $\nu$ predicted by the Transformer. For each set of single-shot measurement of eigenvalues obtained in long times, the Transformer predicts a value of $\nu$. The average $\nu$ from three set of measurements is shown, and the error bar denotes standard deviation. $\mathbf{c1}$-$\mathbf{d2}$ Attention weights projected onto the preprocessed experimental data, shown in $\mathbf{c1}$--$\mathbf{c2}$ the complex energy plane and $\mathbf{d1}$--$\mathbf{d2}$ as a function of $\theta$. }\label{Fig6}
	\end{center} 
\end{figure}

Note that the BEC is in the regime $n\lambda^3\gg 1$, where it exhibits collective response to resonant light~\cite{You1994,Pellegrino2014,Jenkins2016,Schilder2020,Wang2024}. This gives rise to a density-dependent dissipation with many-body nature (see Methods and Supplementary Sec.~III), in contrast to the single-particle dissipation in previous experiments~\cite{Wang2021,Patil2022,Zhang2023,Cao2023,Wu2023,Rao2024}. \color{black}To extract the resulting spectral properties, we monitor the system's decay dynamics over $\sim 0.15$ ms. Initially, a BEC with $N_0\sim 2.5 \times10^4$ $^{87}\textrm{Rb}$ atoms is prepared in $|1\rangle$ at the temperature $\sim 50$ nK (Fig.~\ref{Fig4}$\mathbf{a}$). We then sweep the bias magnetic field to $16$ G over a $100$-ms and wait $100$-ms for stabilization. Subsequently, we switch on both $\Omega$ and the dissipation laser. After evolution time $t$, we measure atomic populations $N_1(t)$ and $N_2(t)$ in the two Zeeman levels $|1\rangle$ and $|2\rangle$, respectively, via Stern-Gerlach absorption imaging; see Figs.~\ref{Fig4}$\mathbf{b1}$-$\mathbf{b2}$. 

At short times, the measured dynamics remains well described by the effective Hamiltonian~(\ref{eq4}) with the constant dissipation rate $\Gamma'$ (solid curves in Figs.~\ref{Fig4}$\mathbf{b1}$-$\mathbf{b2}$). The corresponding calculations $N_1(t)=e^{-\Gamma't}|\cos(\delta t)+i\frac{\Delta'-i\Gamma'}{2\delta}\sin(\delta t)|^2$ and $N_2(t)/N_0=\frac{\Omega^{2}}{|\delta|^2} e^{-\Gamma^\prime t}|\sin(\delta t)|^2$ explain the initial decaying oscillation. At longer times, however, deviations emerge between the experiment and simulation using $H_e(\Gamma')$, especially for small detuning (see Supplementary Information). This discrepancy stems from the density-dependent dissipation: As the condensate dilutes over time from atomic loss, the dissipation rate at long times may differ considerably from its initial value. To effectively model the full dynamics, we adopt a minimal, two-parameter fitting approach (Methods). Namely, we use two Hamiltonians of the form in Eq.~(\ref{eq4}) but distinct dissipation rates: $H_e(\Gamma_1)$ for $0\le t<t_m$ and $H_e(\Gamma_2)$ for  $t\ge t_m$, where ${\Gamma}_{1,2}$ and $t_m$ are obtained from the best fit to the data. As shown in Figs.~\ref{Fig4}$\mathbf{b1}$-$\mathbf{b2}$ and detailed in Supplementary Sec.~III, this approach yields good agreement with the experiment; moreover, $\Gamma_1\approx\Gamma'$ and hence follows $\Gamma_1(\theta)/\Omega\approx -2J_3\sin (m\theta)+ \gamma$, whereas $\Gamma_2<\Gamma_1$ and $\Gamma_2(\theta)/\Omega$ deviates markedly from this relation. 

The eigenvalues of $H_e(\Gamma_1)$ and $H_e(\Gamma_2)$ thus allow us to reliably construct the instantaneous complex-energy braids in the short- and long-time limits, respectively. \color{black}In Figs.~\ref{Fig4}$\mathbf{c1}$-$\mathbf{c3}$, we collect the eigenvalues of $H_e(\Gamma_1)$ to trace their trajectories in $(\mathrm{Re}(E), \mathrm{Im}(E), \theta)$ space; each data is the average of $3$ single-shot measurements. Since $\Gamma_1\approx \Gamma'$, the resulting braids closely match the ideal forms from Eq.~(\ref{Etheta}), displaying the unlink, unknot, and Hopf link, respectively. In the long-time limit, however, spectra of $H_e(\Gamma_2)$ differ substantially from short-time configurations due to $\Gamma_2<\Gamma_1$ as we shall show below. 

Our goal is to identify not only the topological invariant of the experimentally constructed energy braids, but also the underlying geometric features\color{black}. We therefore feed the fitted spectral data into the trained Transformer. Due to the finite number of $\theta$-points attainable from the limited resolution in varying $\Delta'$ and $\Gamma'$, the raw energy data are first preprocessed using the cubic spline interpolation (Methods) to standardize the sequence length to the Transformer’s required input dimension while enhancing smoothness. We then sequentially and uniformly sample $500$ points along the interpolated curve (e.g., solid curves in Figs.~\ref{Fig5}$\mathbf{a1}$-$\mathbf{a3}$), including all original experimental results. The resulting processed complex energies $E_\pm (\theta)$ are used as the input to the Transformer. 

Figures~\ref{Fig5}$\mathbf{b}$-$\mathbf{d3}$ summarize typical ML results obtained from processed single-shot measurement data in the short-time regime. The Transformer accurately predicts braiding degree $\nu$ for various experimental braids. Even at long times, when the energy braids deviate substantially from their initial ideal configurations, the Transformer correctly identifies their topology. For instance, initially topologically nontrivial braids (unknot or Hopf link, Figs.~\ref{Fig5}$\mathbf{a2}$-$\mathbf{a3}$) eventually evolve into the unlink structure characterized by $\nu=0$ (Figs.~\ref{Fig6}$\mathbf{a1}$-$\mathbf{c2}$), indicating a topological phase transition driven by the dynamically reduced dissipation rate. Moreover, in both the short- and long-time regimes, the attention maps (Figs.~\ref{Fig5}$\mathbf{d1}$-$\mathbf{d3}$ and~\ref{Fig6}$\mathbf{d1}$-$\mathbf{d2}$) consistently highlight those regions near band-crossings as important for the topological classification; more details can be found in Supplementary Sec.~V. These results show that the Transformer, though trained exclusively on the symmetric spectra derived numerically from a nearly ideal Hamiltonian, generalizes effectively to an experimental system that breaks sublattice symmetry and hosts time-varying energy braids induced by density-dependent dissipation. 

In both simulated tests (Fig.~\ref{Fig3}) and experimental demonstrations (Figs.~\ref{Fig5}-\ref{Fig6}), we observe that attention peaks may slightly shift from the exact band-crossing positions. This is likely linked to the training objective of the Transformer: it is solely optimized for predicting topological invariant with high confidence, not for precise detection of band-crossing point. During training, the self-attention mechanism learns to assign weight to those momentum‑energy regions that are most informative for a correct topological decision. Since topological invariants are inherently nonlocal, the model naturally integrates information from adjacent spectral points, distributing attention weights over a neighborhood surrounding a band-crossing rather than precisely pinpointing its location. This may cause the shift of attention peaks, and their relative prominence can depend on the specific geometry of the braid. Thus the attention map serves more as a semi‑quantitative diagnostic tool that explicitly guides our insight toward spectral regions containing topologically relevant geometric features. We emphasize that this dual capability - accurate topological classification combined with autonomous identification of essential parts in the braids - is a crucial advantage of the Transformer over conventional methods. Achieving more accurate local feature identification while preserving the prediction precision of nonlocal topological invariant is a challenge beyond the current architecture and will be addressed in future work.

\bigskip

\noindent\textbf{Discussion}  

In summary, we experimentally realize and characterize topological complex-energy braiding in an atomic simulator based on dissipative BECs. By harnessing the Transformer's self-attention mechanism, we achieve a simultaneous extraction of not only topological invariants of complex-energy braiding but also their governing geometric features - specifically, the band-crossing with extreme phase gradients. That is, the attention weights in the input data correctly spot the physically relevant regions underlying braiding topology. Moreover, in our experiment, tunable dissipation has a many-body nature, and features density dependence. This feature potentially makes dissipative BECs under resonant light a unique
platform in the study of non-Hermitian physics, yet at the same time complicates experimental detections of topology-geometry interplay. Remarkably, the Transformer, trained exclusively on simulated, symmetric spectra from an ideal single-particle Hamiltonian, demonstrates robust generalization to experiment. This capability of directly capturing the topology-geometry interplay with ML opens up many fascinating opportunities, for instance, to investigate the interplay between topology and quantum geometry, which has attracted significant recent attention, yet experimental studies remain challenging. In the future, it is interesting to explore a comprehensive microscopic description of the full dissipative dynamics of a BEC and real-time topological
diagnosis. Another compelling direction is to extend this ML paradigm to explore topological phases in many-body regimes. More broadly, our work is relevant to the development of interpretable ML approaches~\cite{Ponte2017,Greitemann2019,Dawid2020,Wetzel2020,Zhang2020,Miles2021} to facilitate experimental studies of exotic topological phases in cold atoms and beyond.\\

Recent work~\cite{Huili2022, Si2023} has demonstrated that ML approaches for phase classifications are vulnerable to adversarial perturbations: adding carefully crafted noises into the original data can cause the classifiers to make wrong predictions. We expect the Transformer to share this vulnerability. Such vulnerability, in fact, can be a generic issue for data-driven classifiers and is often linked to the geometry of the loss landscape~\cite{Pierre2010}, i.e., models converging to flat minima typically exhibit stronger robustness compared to those converging to sharp minima. Adversarial training~\cite{Yuan2019,Chakraborty2021} and refined optimization strategies can help mitigate this issue.

\bigskip

\noindent\textbf{Methods}  

\noindent\textbf{Self-attention mechanism of the Transformer and generation of heatmap} 

The output of the attention mechanism~\cite{Vaswani2017} is given by
\begin{equation}\label{attention}
\textrm{Attention}(Q,K,V) = \textrm{softmax}\left(\frac{Q K^{{T}}}{\sqrt{d_{k}}}\right){V}\equiv SV,
\end{equation}
where $d_{k}$ is the dimension of expanded input, query $Q$, key $K$, and value $V$ vectors result from linear transformations of the input using distinct weight matrices $W_{q}$, $W_{k}$, and $W_{v}$. In Eq.~(\ref{attention}), attention weights are generated from the softmax function of scaled similarities between queries and keys. These weights are then used to calculate a weighted sum of the value vectors (see Supplementary Sec.~I). To enhance feature extraction, the multi-head attention mechanism projects query ($Q$), key ($K$), and value ($V$) vectors into multiple subspaces, performs self-attention independently per head, and generates the outputs. 

The attention weight matrix $S$ encodes the learned correlations between all pairs of momentum points in the BZ. Each element of matrix $S$ quantifies the relevance of  $k_j$ to $k_i$. For geometric visualization and physical interpretation, we extract attention importance scores by computing column-wise averages of the attention weight matrix, yielding a weight score vector of dimensionality $1 \times 500$. The weights are then projected onto the energy band space to generate the corresponding heatmap in Fig.~\ref{Fig3}, where weights quantitatively encode feature important for each topological invariant.

\bigskip
\noindent\textbf{Test data for the Transformer} 

To validate the performance of the trained Transformer, we use the energy spectrum distinct from Eq.~(\ref{eq1}) as the test data. Specifically, we consider
\begin{equation}\label{Etest}
E_\pm (k)=\alpha\left(\frac{\Delta^{\prime\prime} - i\Gamma^{\prime\prime}}{2}\right)\pm \sqrt{ \left( \frac{\Delta^{\prime\prime} - i\Gamma^{\prime\prime}}{2} \right)^2+\Omega^2},
\end{equation}
Here, $\Delta^{\prime\prime}/\Omega=2J_1 \cos k+2J_2\cos m k+2J_{4} \cos 2 k$ and $\Gamma^{\prime\prime}/\Omega= -2J_3 \sin m k+ \gamma$, in which the parameters $J_i$, $\gamma$ (i=1, 2, 3, 4) are arbitrarily selected, $m=1, 2..., 6$, and $k \in [0, 2\pi]$. Moreover, $\alpha=0$ or $\alpha=1$ are used. The corresponding test results are illustrated in Fig.~\ref{Fig3} and Supplementary Sec.~II.

\bigskip

\noindent\textbf{Measurement of dissipation rate}

To measure the dissipation rate $\Gamma'$ induced in $\left | F=2,m_F=0  \right \rangle $ (Fig.~\ref{Fig4}$\mathbf{a}$), we use MW $\pi$ pulse to transfer the BEC from $\left | F=1,m_F=-1  \right \rangle $ to $\left | F=2,m_F=0  \right \rangle $. Subsequently, the MW field is switched off while the resonant laser is switched on, and we measure the time-dependent population $N'(t)$ in $\left | F=2,m_F=0  \right \rangle $. By fitting $N'(t)=N_0e^{-2\Gamma' t}$ to the decay dynamics at sufficiently short times, we obtain $\Gamma'$. In Supplementary Sec.~III and Fig.~11, we present the measured $\Gamma'$ as a function of the power $P$ of the dissipation laser for various atomic densities $n$. We observe enhanced dissipation for increased density, especially at higher $P$. Moreover, for $n \sim 5.14\times 10^{13}\textrm{cm}^{-3}$ in our experiment, $\Gamma'=0.68P$ (Supplementary Fig.~11). Thus, adjusting $P$ allows to sample $\Gamma'$ along the relation $\Gamma'/\Omega= -2J_3 \sin (m\theta) + \gamma$ with given $J_3,m,\gamma$, where each $\theta$-point (Supplementary Fig.~13) maps to a value of $P$. 

Due to the density-dependent dissipation of condensate, although $H_e(\Gamma')$ remains a good description for system dynamics at short times, suppressed decay dynamics may occur at long times. To model the full dynamics over a relatively long time scale of $\sim 0.15$ ms, we exploit a minimal strategy: the two-parameter fit~\cite{Wang2024}. Specifically, we use two Hamiltonians of the form (\ref{eq4}), $H_e({\Gamma}_{1})$ and $H_e({\Gamma}_{2})$, to model the evolution at times $0\le t<t_m$ and $t\ge t_m$, respectively, with  ${\Gamma}_{1}$, ${\Gamma}_{2}$ and $t_m$ to be determined from the best fit to the data. In fitting, we take the initial value of $\Gamma_1$ to be $\sim\Gamma'$, and the initial $\Gamma_{2}$ to be $\sim\Gamma'/2$; for the initial value of $t_m$, we take it to be where the data deviate from the simulation with $H_e(\Gamma')$. Starting from these initial values, we iteratively improve the fitting parameters until the difference of the calculated and the measured dynamics are minimized in the least-square sense. As detailed in Supplementary Sec.~III, the two-parameter fit generally yields good agreement with the experimental data. The collection of fitted $\Gamma_1$ and $\Gamma_2$ are summarized in Supplementary Fig.~13. 
\bigskip

\noindent\textbf{Measurement of energy eigenvalues and preparation of input data} 

By diagonalizing the Hamiltonian $H_e$ of the form in Eq.~(\ref{eq4}) with $\Gamma_{1(2)}$, we obtain the experimental spectra corresponding to the short-time (long-time) limit. 

Subsequently, we process the experimentally measured data in two steps to generate the input data of $E_\pm(\theta)$ for the trained Transformer. First, we employ a standard interpolation procedure based on cubic spline functions to convert the discrete, irregularly spaced experimental data into a continuous curve. Rather than fitting a single high-degree polynomial to all data, cubic spline interpolation - a widely used method in data analysis - constructs a smooth curve that passes exactly through the given discrete data points. This approach is robust, numerically stable, and capable of preserving the shape of the underlying data. Specifically here, let the initial experimental data be denoted by $[\textrm{Re}E_\pm(\theta_n), \textrm{Im}E_\pm(\theta_n)]$, each associated with a parameter point $\theta_n$ (with $\theta_n$ ordered by increasing magnitudes over a period $[0,2\pi]$.  The cubic spline function is used to fit the relation between $\textrm{Re}E_\pm$ and $\theta_n$, as well as between $\textrm{Im}E_\pm$ and $\theta_n$. This result in curves in Figs.~\ref{Fig5} and \ref{Fig6}. Further details can be found in the Supplementary Sec.~IV. Second, after obtaining the interpolated curve, $500$ complex-energy points are sequentially sampled at positions $\theta_n = 2\pi n/500$ ($n = 0, ..., 499$) along the curve, including all original experimental data. These complex-energies are then used to generate the input data for the Transformer. 

\bigskip
\noindent\textbf{Data Availability}  

\indent The experimental data generated in Figs.~4-6 have been deposited in the Zenodo database under accession code [\url{https://doi.org/10.5281/zenodo.18538248}].  

\bigskip
\noindent\textbf{Code Availability}  

 \indent The codes that support the findings of this study are available in the Code Ocean database under accession code [\url{https://codeocean.com/capsule/1694502/tree}].

\bigskip
\noindent\textbf{Acknowledgments}

This research is funded by the Key Project of the National Natural Science Foundation of China Joint Funds (No.~U25A20197),  the National Key Research and Development Program of China (No.~2022YFA1404201, No.~2022YFA1203903, No.~2022YFA1404003), the National Natural Science Foundation of China (No.~12374246, No.~12274272, No.~12504308, No.~62432006), and 111 Project (grant No.~D18001). Y. H. acknowledges support by Beijing National Laboratory for Condensed Matter Physics (No. 2023BNLCMPKF001). L.B. acknowledges support by the Fundamental Research Program of Shanxi Province (No.~202303021223004). 

\bigskip

\noindent\textbf{Author contributions}

Y.H., L.B., and Y.Z. jointly conceived the research idea and supervised the entire study. Y.Y. and X.Z. developed and implemented the machine learning. N.L. and Y.H. performed the theoretical calculation and analysis. C.W., Z.J. and Y.Z. performed the experimental measurements related to the Bose-Einstein condensate. N.L., Y.Y. and Z.F. contributed to the processing of experimental data and validation of machine learning performance. Y.Y., N.L., X.Z. and Y.H. wrote the manuscript, Z.J., L.X., and S.J., provided critical guidance throughout the project.\\

\noindent\textbf{Competing interests}

The authors declare no competing interests.\\

\end{document}